# Transport Dynamics of Water Molecules Confined between Lipid Membranes

*Minho Lee[1,2], Euihyun Lee[3], Ji-Hyun Kim[1,2], Hyonseok Hwang[6], Minhaeng Cho[4,5]\*, and Jaeyoung Sung[1,2]\**

[1] Creative Research Initiative Center for Chemical Dynamics in Living Cells, Chung-Ang University, Seoul 06974, Republic of Korea

[2] Department of Chemistry, Chung-Ang University, Seoul 06974, Republic of Korea

[3] Department of Chemistry, The University of Texas at Austin, TX 78757, USA

[4] Center for Molecular Spectroscopy and Dynamics, Institute for Basic Science (IBS), Seoul 02841, Republic of Korea

[5] Department of Chemistry, Korea University, Seoul 02841, Republic of Korea

[6] Department of Chemistry, Institute for Molecular Science and Fusion Technology, Kangwon National University, Chuncheon, Gangwon-do 24341, Republic of Korea

\*Correspondence and requests for materials should be addressed to J.S. (email: jaeyoung@cau.ac.kr) and M.C. (email: mcho@korea.ac.kr).

ABSTRACT

Water molecules confined between biological membranes exhibit a distinctive non-Gaussian displacement distribution, far different from bulk water. Here, we introduce a new transport equation for water molecules in the intermembrane space, quantitatively explaining molecular dynamics simulation results. We find that the unique transport dynamics of water molecules stems from the lateral diffusion coefficient fluctuation caused by their longitudinal motion in the direction perpendicular to the membranes. We also identify an interfacial region where water possesses distinct physical properties, unaffected by changes in the intermembrane separation.

**TOC GRAPHICS**

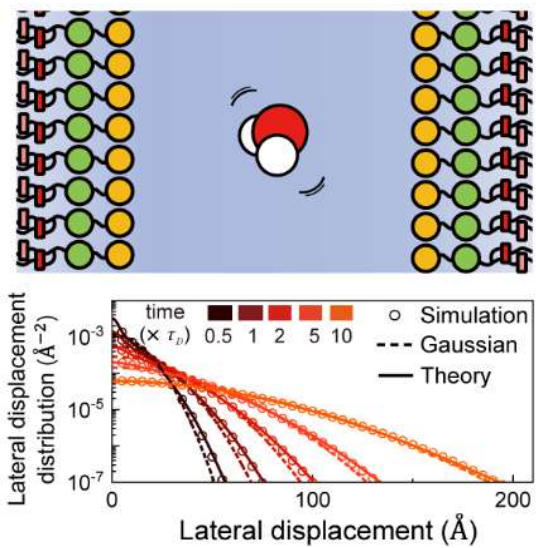

Water confined in biological nanospaces, such as the inter-membrane regions of mitochondria, synaptic clefts, and endoplasmic reticula, is a crucial element for cell function[1–3]. Extensive research has investigated the structure and dynamics of water molecules in the vicinity of phosphatidylcholine (PC) bilayers, a major component of biological membranes. These works employed various methods including IR pump-probe spectroscopy[4–6], heterodyne-detected vibration sum frequency generation (HD-VSFG)[7,8], nuclear magnetic resonance (NMR)[9], microfluidics[10–12], and molecular dynamics (MD) simulation[13–21]. It is now established that the motility of nanoconfined water increases with its distance from the membrane center, with which the major functional groups of PC phospholipids interacting with water molecules change[4,9,14,17,18]. Moreover, it has been observed that dynamic motility fluctuations lead to Fickian-yet-non-Gaussian diffusion in complex fluids[22]. Despite these studies, however, a quantitative understanding of the time-dependent displacement distribution of nanoconfined water molecules has not yet been achieved.

To address this issue, we introduce a transport equation describing the thermal motion of molecules confined between two planar surfaces. Using MD simulations, we investigate the time-dependent displacement distribution of water molecules nanoconfined between two lipid membranes composed of 1,2-dimyristoyl-sn-glycero-3-phosphocholine (DMPC) lipids. The solution derived from our transport equation provides a quantitative explanation of the MD simulation results for the mean square displacement, the non-Gaussian parameter, and the displacement distribution. Our analysis shows that nanoconfined water exhibits a super-Gaussian lateral displacement distribution originating from dynamic fluctuations of the lateral diffusion coefficient due to its coupling to water motion in the longitudinal direction. The time-dependent deviation of this lateral displacement distribution from Gaussian is found to be strongly influenced by the intermembrane separation. We also identify an interfacial region where water molecules have structure and dynamics that are distinct from bulk water and robust with respect to changes in intermembrane separation.

The essential assumption underlying our transport equation is that the most important variable that affects the lateral thermal motion of an interfacial water molecule is the distance, $z$, between the water molecule and the center of the membrane. This is a legitimate assumption because the microscopic environment interacting with a water molecule, including the functional groups of lipid molecules and hydrogen bond network, drastically changes with $z$[9,17,18]. Under this assumption, we obtain the following transport equation governing transport dynamics of interfacial water, starting from a general model of thermal motion coupled to environmental variables[23]:

$$\hat{\dot{p}}(\mathbf{r}_\parallel, z, s) = \hat{\mathcal{D}}_\parallel(z, s)\nabla_\parallel^2 \hat{p}(\mathbf{r}_\parallel, z, s) + L(z)\hat{p}(\mathbf{r}_\parallel, z, s), \qquad (1)$$

where $\hat{p}(\mathbf{r}_\parallel, z, s)$ denotes the Laplace transform of the joint probability density, $p(\mathbf{r}_\parallel, z, t)$, that a water molecule is located at lateral position $\mathbf{r}_\parallel$ ($= (x, y)$) and the distance between the water molecule and the center of the membrane is $z$ at time $t$. $\hat{f}(s)$ and $\hat{\dot{f}}(s)$ denote the Laplace transform of $f(t)$ and $\partial_t f(t)$, i.e., $\hat{f}(s) = \int_0^\infty dt e^{-st} f(t)$ and $\hat{\dot{f}}(s) = \int_0^\infty dt e^{-st} \partial_t f(t)$. $\nabla_\parallel^2$ denotes the Laplacian in the two-dimensional space parallel to the lipid membrane. In eq 1, $\hat{\mathcal{D}}_\parallel(z, s)$ represents the lateral diffusion kernel dependent on $z$. Its small-$s$ limit, $\hat{\mathcal{D}}_\parallel(z, 0)$, serves as the lateral diffusion coefficient, $D_\parallel(z)$, of water molecules separated by $z$ from the center of the membrane. In eq 1, $L(z)$ denotes a mathematical operator describing the transport dynamics of the water molecules in a direction perpendicular to the membrane surface.

We then investigate the mean square displacement (MSD), $\Delta_2(t)$, and non-Gaussian parameters (NGP), $\alpha_2(t)\left[\equiv \Delta_4(t)/(2\Delta_2(t)^2)-1\right]$, of the lateral displacement distribution of the water molecules. Here, $\Delta_2(t)$ and $\Delta_4(t)$ denote the second and fourth moments of the time-dependent distribution of the water displacement, $\Delta \mathbf{r}_\|(t)\left[\equiv \mathbf{r}_\|(t)-\mathbf{r}_\|(0)\right]$, in the lateral direction. The NGP vanishes when the displacement distribution is Gaussian[24,25]. From eq 1, analytic expressions of the $\Delta_2(t)$ and $\Delta_4(t)$ can be obtained as

$$\hat{\Delta}_2(s) = \frac{4}{s^2}\langle \hat{\mathcal{D}}_\|(s)\rangle, \qquad (2a)$$

$$\hat{\Delta}_4(s) = 4s\hat{\Delta}_2(s)^2\left[1+s\hat{C}_\mathcal{D}(s)\right]. \qquad (2b)$$

In eq 2a, $\langle \mathcal{D}_\|(t)\rangle$ denotes the mean diffusion kernel of water molecules. $\langle Q \rangle$ designates the average of quantity $Q(z)$ over the equilibrium distribution, $P_{eq}(z)$, of $z$. $\langle \mathcal{D}_\|(t)\rangle$ is the same as the lateral velocity autocorrelation function (VAF) of water molecules divided by $2^{23}$. In eq 2b, $C_\mathcal{D}(t)$ denotes the lateral diffusion kernel correlation (DKC) defined by

$$\hat{C}_\mathcal{D}(s) = \int_0^\ell dz \int_0^\ell dz_0 \frac{\delta \hat{\mathcal{D}}_\|(z,s)}{\langle \hat{\mathcal{D}}_\|(s)\rangle} \hat{G}(z,s|z_0) \frac{\delta \hat{\mathcal{D}}_\|(z_0,s)}{\langle \hat{\mathcal{D}}_\|(s)\rangle} P_{eq}(z_0), \qquad (3)$$

where 0 and $\ell$ denote the center positions of the two membranes confining water molecules (see Figure 1). In eq 3, $\delta \hat{\mathcal{D}}_\|(z,s)$ and $G(z,t|z_0)$ denote, respectively, $\hat{\mathcal{D}}_\|(z,s)-\langle \hat{\mathcal{D}}_\|(s)\rangle$ and the Green's function, or the conditional probability that a water molecule initially located at $z_0$ is found at $z$ at time $t$, defined by $\partial_t G(z,t|z_0) = L(z)G(z,t|z_0)$ with the initial condition, $G(z,0|z_0) = \delta(z-z_0)$. Equation 2 enables us to extract the time profile of the DKC from the MSD

and the NGP or the first two nonvanishing moments of the displacement distribution (see Figure S3a in Supporting Information).

At the onset of Fickian diffusion, the NGP reaches its maximum value (Figure 2). Beyond this NGP peak time, the MSD of water molecules linearly increases with time, which results because the VAF or $\langle \mathcal{D}_\parallel(t) \rangle$ is negligibly small after the NGP peak time, i.e., $\Delta_2(t) = 4\int_0^t d\tau (t-\tau) \langle \mathcal{D}_\parallel(\tau) \rangle \cong 4t \int_0^\infty \langle \mathcal{D}_\parallel(\tau) \rangle$. At time scales longer than the NGP peak time, $L(z)$ in eq 1 can be approximated by the following Smoluchowski operator, i.e., $L(z) \cong L_{\text{SM}}(z) = \partial_z [D_\perp(z)(\partial_z + \partial_z \beta U(z))]$. Here, $D_\perp(z)$ and $\beta U(z)$, respectively, denote the z-dependent diffusion coefficient associated with the thermal motion of water molecules in the direction perpendicular to the membrane and the thermal energy-scaled potential of mean force with $\beta = 1/k_B T$. $k_B$ and $T$ denote the Boltzmann constant and temperature, respectively. After the onset of the Fickian diffusion, the MSD and NGP of the lateral diffusion of water molecules assume the following analytic forms:

$$\Delta_2(t) \cong 4 \langle D_\parallel \rangle t, \tag{4a}$$

$$\alpha_2(t) \cong \frac{2\eta_D^2}{t^2} \int_0^t dt' (t-t') \phi_D(t'). \tag{4b}$$

In eq 4, $\langle D_\parallel \rangle$ and $\eta_D^2 \left[ = \langle \delta D_\parallel^2 \rangle / \langle D_\parallel \rangle^2 \right]$ denote, respectively, the mean diffusion coefficient defined by $\langle D_\parallel \rangle = \int_0^\infty \langle \mathcal{D}_\parallel(\tau) \rangle$ and the relative variance of the z-dependent lateral diffusion coefficient, with $\delta f$ denoting the deviation of a quantity $f$ from its mean, i.e., $\delta f(z) = f(z) - \langle f \rangle$. $\phi_D(t)$ denotes the normalized time-correlation function (TCF) of the lateral diffusion coefficient fluctuation, i.e.,

$$\phi_D(t) = \frac{\langle \delta D_\parallel(t)\delta D_\parallel(0)\rangle}{\langle \delta D_\parallel^2\rangle} = \langle \delta D_\parallel^2\rangle^{-1}\int_0^\ell dz \int_0^\ell dz_0 \delta D_\parallel(z) G_{SM}(z,t|z_0)\delta D_\parallel(z_0) P_{eq}(z_0), \quad (5)$$

where $G_{SM}(z,t|z_0)$ designates Green's function of Smoluchowski equation governing the thermal motion of water molecules in the direction perpendicular to the membranes, i.e., $\partial_t G_{SM}(z,t|z_0) = L_{SM}(z) G_{SM}(z,t|z_0)$, with the initial condition, $G_{SM}(z,0|z_0) = \delta(z-z_0)$.

Equations 4b and 5 indicate that, for nanoconfined water molecules, the non-Gaussian diffusion in the lateral direction originates from fluctuation of the lateral diffusion coefficient coupled to water motion in the longitudinal direction. We note here that the DKC defined in eq 3 reduces to $\eta_D^2 \phi_D(t)$ at long times where the MSD linearly increases with time; the long-time profile of $C_D(t)$ extracted from the MSD and NGP time profiles of our MD simulation results is in quantitative agreement with our theoretical result for $\eta_D^2 \phi_D(t)$ calculated using eq 5. This agreement between simulation and theory supports the validity of our assumption underlying eq 1 that the dynamic fluctuation in the lateral diffusion coefficient of water molecules primarily originates from its coupling to the thermal motion of water molecules in the longitudinal direction.

We also performed MD simulation study on the thermal motion of water molecules in the intermembrane space, for a system of SPC/E water molecules confined between two lipid bilayers, each composed of 128 DMPC molecules, systematically changing the ratio of the number of water molecules to the number of lipid molecules and the distance $\ell$ between the two membrane centers (Figure 1). We employed the AMBER lipid 14 force field for simulation of the lipid molecules[26] and imposed periodic boundary conditions on the system. We also investigated the transport dynamics of a pure, bulk water system, using the MD simulation of 5000 SPC/E water molecules, and compared it to the transport dynamics of the nanoconfined water molecules. Our MD simulation was conducted for a system at a temperature of 318 K in the *NVT* ensemble. Additional information about the simulations can be found in the Supporting Information.

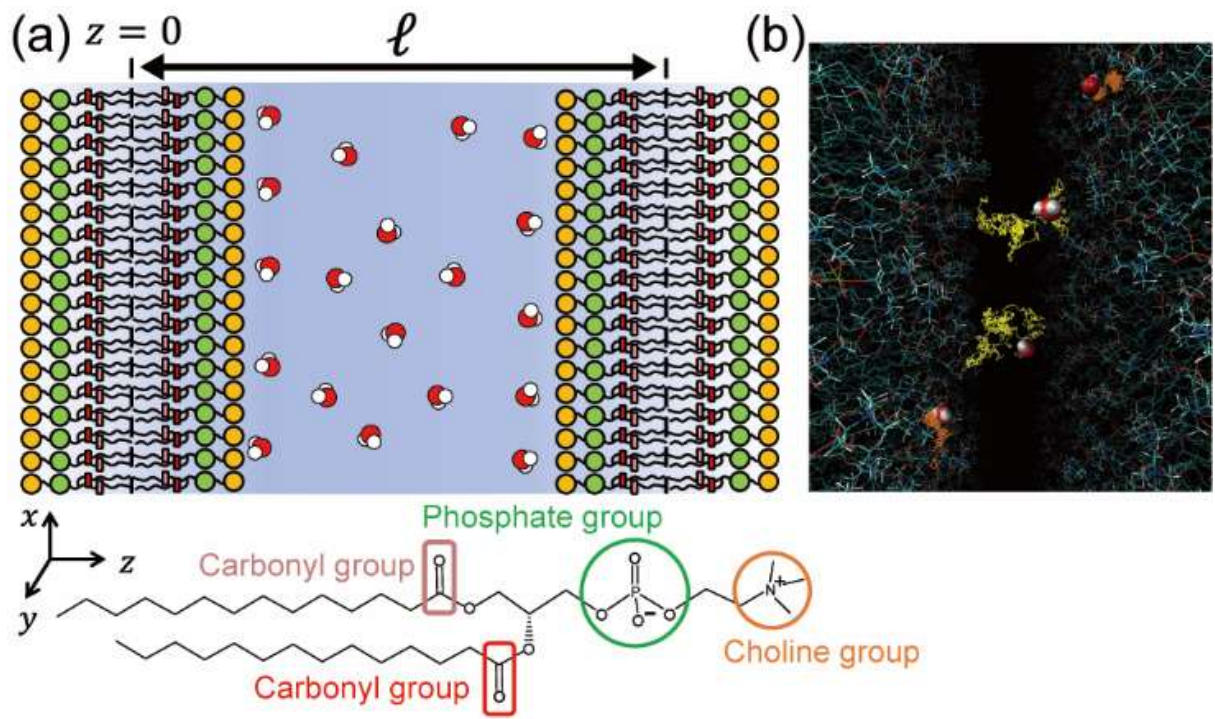

**Figure 1.** (a) Schematic representation of our MD simulation system: water molecules confined between two DMPC lipid bilayers. The simulation was performed for the system with three different intermembrane separations, i.e., $\ell$ = 53.2 Å, 61.8 Å, and 80.6 Å. (b) Representative trajectories of water molecules undergoing thermal motion between the lipid bilayers for 50 ps: (yellow lines) the trajectories of water molecules freely moving near the center of the intermembrane space; (orange lines) trajectories of water molecules strongly interacting with the lipid head groups of DMPC.

From the MD simulation trajectories of water molecules, we obtained the MSD and NGP of the lateral water displacement for each system (Figure 2). The time profile of the MSD obtained from the MD simulation exhibits a dynamic transition behavior that is dependent on the separation between the two lipid membranes (Figure 2a). The time profile of the MSD shows the transition from an initial ballistic motion ($\Delta_2(t) \sim t^2$) to terminal Fickian diffusion ($\Delta_2(t) \sim t^1$), with intermediate subdiffusion ($\Delta_2(t) \sim t^\alpha$ with $0 < \alpha < 1$). The short-time ballistic behavior of the MSD shows little variation with changes in $\ell$ and quadratically increases with time, i.e., $\Delta_2(t) = 2k_B T t^2/M$ [23]. Here, $M$ denotes the mass of a water molecule. However, the intermediate subdiffusive regime becomes more pronounced, and the value of long-time lateral diffusion coefficient gets smaller as the separation, $\ell$, between lipid bilayers decreases. These findings align with previous studies[10,12,16,21]. The time profiles of the MSD could be quantitatively explained by using the analytic formula for the MSD of a bead in a Gaussian polymer (see Figure 2a), which is decomposable into an unbound-mode and multiple bound-mode terms[23]. For bulk water at room temperature, the bound-mode terms, which cause the intermediate subdiffusion, are negligible. However, for intermembrane water, the contribution from bound-modes increases as the separation, $\ell$, between the membranes decreases. This indicates that the bound mode terms result from interfacial water molecules, which are transiently trapped by lipid membranes.

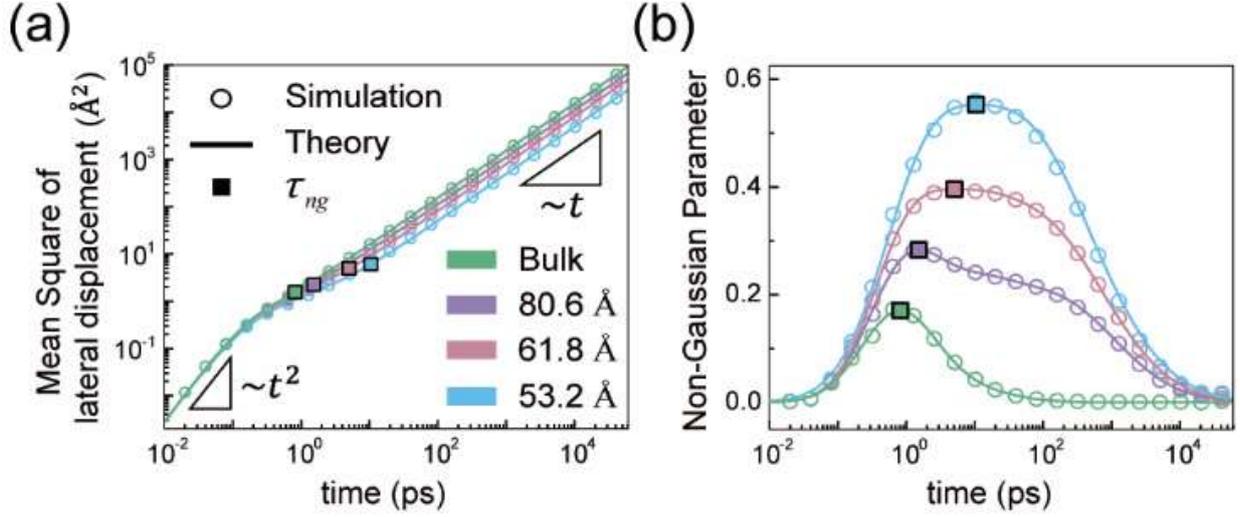

**Figure 2.** (a) Mean square displacement (MSD) and (b) non-Gaussian parameter (NGP) associated with the lateral displacement of water molecules for systems with various intermembrane separations: (circles) simulation results; (solid lines) eq S19 for MSD and eq S21 for NGP; (square) the NGP peak time ($\tau_{ng}$)

Our MD simulation study reveals that both the NGP peak time and the peak height increase as the separation, $\ell$, between the membranes decreases (Figure 2b). A longer peak time and a greater peak height of the NGP signify, respectively, prolonged trapping of water molecules[23,27] and increased fluctuation in the lateral diffusion coefficient (see eq 4b). These phenomena can be attributed to the attractive interactions between interfacial water molecules and the functional groups in the lipid molecules. As $\ell$ decreases, the proportion of the trapped interfacial water molecules grows, leading to a decrease in the mean lateral diffusion coefficient and an increase in the variance of the diffusion coefficient.

At times longer than the NGP peak time, the nanoconfined molecules undergo Fickian yet non-Gaussian diffusion, with the NGP value decreasing with time. The nanoconfined water molecules have a far greater NGP value than bulk water molecules. The NGP of bulk water molecules becomes negligible at times longer than 10 ps; in contrast, the NGP of the nanoconfined water molecules does not vanish at times longer than 10 ns. The NGP shows a strongly non-exponential relaxation dynamics, whose time profile can be quantitatively explained by eqs 4b and 5 as shown later in this Letter.

We then identify the interfacial region where water molecules directly interact with the lipid head groups. For this purpose, we investigate the structure and dynamics of water molecules near the lipid-bilayers using the MD simulations. Figures 3a and 3b show the $z$-dependence of the density profile, $\rho(z)$, and the dipole orientation profile, $\langle\cos\theta\rangle_z$, of the nanoconfined water. Here, $\theta$ and bracket $\langle\cdots\rangle_z$ denote the angle between the water dipole moment and the $z$-axis and the average over water molecules located within an interval ($z$–0.25 Å, $z$+0.25 Å), respectively. As shown in Figure 3a, the water molecules in close vicinity of the membrane exhibit a lower density compared to the density, $\rho_{bulk}^{45°C} = 0.99$ g·cm$^{-3}$, of bulk water[28]. The dipole orientation profile, $\langle\cos\theta\rangle_z$, does not vanish for water molecules in the vicinity of the functional groups in the lipid molecules. This results from the non-isotropic interactions between the water molecules and those functional groups, whose positions along the $z$-direction are shown in Figure 3a. At positions near $z \cong 20$ Å, $\langle\cos\theta\rangle_z$ has a negative value, while at positions near $z \cong \ell - 20$ Å, it has a positive value, indicating that the water dipole moment tends to point toward phosphate and carbonyl groups rather than choline groups[7,8]. This is because water molecules in the vicinity of phosphate and carbonyl groups experience strong restrictions on their orientations due to the hydrogen bonding with these groups, whereas those near choline groups exhibit broadly distributed angle distribution, forming the clathrate-like hydration shell around the choline groups[4,13].

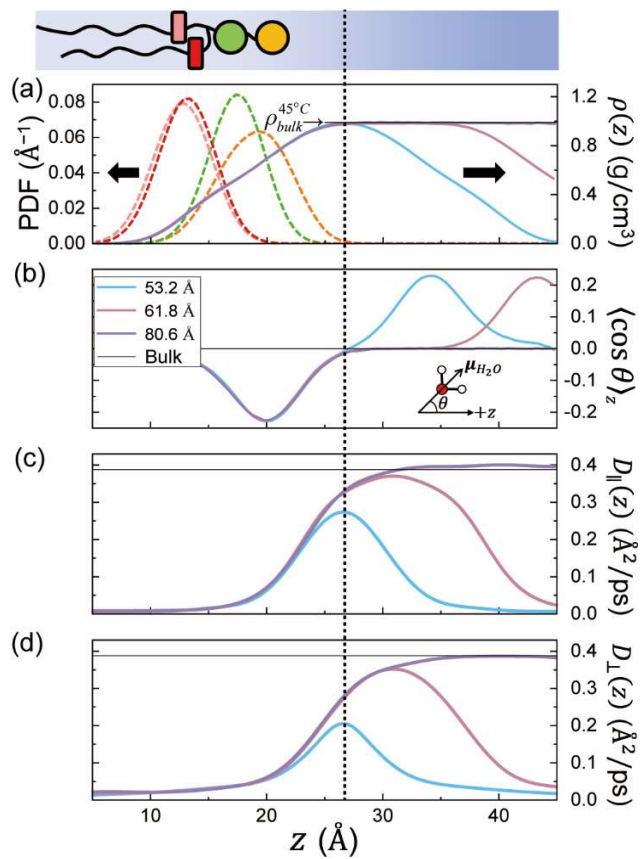

**Figure 3.** Dependence of structural and dynamical properties of the water molecules on the distance $z$ from the center of the lipid membrane in the left. z-dependent profile of (a) mass density, $\rho(z)$, (b) orientation $\langle \cos\theta \rangle_z$, (c) lateral diffusion coefficient, $D_\parallel(z)$, and (d) longitudinal diffusion coefficient, $D_\perp(z)$, of water molecules. $\theta$ designates the angle between the water moment and the longitudinal axis. (dotted line) the position of the boundary, $z_c = 26.6$ Å, between interfacial water region and bulk-like water region. (dashed lines) probability distribution of the positions of various functional groups in the DMPC: (orange) the nitrogen atom in the choline group; (green) phosphorus atom in the phosphate group; (red and pink) the oxygen atoms in the two carbonyl groups. The probability distribution of these functional groups as well as the z-dependent profiles of $\rho(z)$ and $\langle\cos\theta\rangle_z$ are largely independent of the intermembrane separation.

Remarkably, the density and dipole orientation profiles show little dependence on $\ell$ for the interfacial water molecules at $z$ smaller than $z_c = 26.6$ Å but recover their respective bulk-limit values in the intermediate region defined by $z_c \leq z \leq \ell - z_c$. Based on these observations, we define $z_c$ as the boundary between the interfacial water region and the bulk-like water region. The value of $z_c$ (= 26.6 Å) estimated from our MD simulation is found to be comparable to the previously reported $z_c$ values, 24 Å and 28 Å, which were estimated by investigating the structural order[15] and the rotational dynamics[19] of water molecules near the lipid membrane.

The diffusion coefficients of water molecules are also strongly dependent on their distance from the membrane. We obtain the z-dependent profiles of $D_\parallel(z)$ and $D_\perp(z)$ (Figures 3c and 3d) using umbrella sampling and mean first passage time analysis (Figures S1 and S2), respectively[14,29]. Within the interfacial water region ($z < z_c$), both the diffusion coefficients increase with $z$, and their z-dependent profiles are also similar across various intermembrane separations. For the system with $\ell = 80.6$ Å, both diffusion coefficients recover the bulk-limit value, $D_{bulk}^{45°C} = 3.88\times10^{-1}$ Å$^2$·ps$^{-1}$, of the SPC/E water at distances greater than $z_c$. However, for the systems with $\ell < 80.6$ Å, the values of $D_\parallel(z)$ and $D_\perp(z)$ are smaller than $D_{bulk}^{45°C}$ across the entire $z$ region. For example, when the value of $\ell$ is 53.2 Å, the maximum values of $D_\parallel(\ell/2)$ and $D_\perp(\ell/2)$ are only about 70% and 50% of $D_{bulk}^{45°C}$, respectively. Noting that the longitudinal position, z, of the central nitrogen atom in the choline group is about 19.3 $\pm$ 4 Å (Figure 3a), these results suggest that the choline group in the lipid molecule retard the translational motion of water molecules located beyond its first hydration shell, which extends about 4.5 Å from the central nitrogen atom in the choline group[13].

The z-dependence profiles of water density $\rho(z)$ and the diffusion coefficients, $D_{\parallel}(z)$ and $D_{\perp}(z)$, are related to the time profile of $\eta_D^2 \phi_D(t)$ and the NGP according to eqs 4b and 5. Equation 4b tells us that, after the onset of Fickian diffusion, the NGP time profile is completely determined by the mean-scaled TCF, $\eta_D^2 \phi_D(t) [= \langle \delta D_{\parallel}(t) \delta D_{\parallel}(0) \rangle / \langle D_{\parallel} \rangle^2]$, of the diffusion coefficient fluctuation. Here, $\eta_D^2 [= \langle D_{\parallel}^2 \rangle / \langle D_{\parallel} \rangle^2 - 1]$ can be calculated from $D_{\parallel}(z)$ and $\rho(z)$ by using $\langle D_{\parallel}^n \rangle = \int_0^{\ell} dz D_{\parallel}^n(z) P_{eq}(z)$ with $P_{eq}(z) = \rho(z) / \int_0^{\ell} dz \rho(z)$. $\phi_D(t)$ can also be calculated from eq 5, where the Green's function $G_{SM}(z,t|z_0)$ is obtained by solving $\partial_t G_{SM}(z,t|z_0) = \partial_z [D_{\perp}(z)(\partial_z + \partial_z \beta U(z))] G_{SM}(z,t|z_0)$ with the initial condition $G_{SM}(z,0|z_0) = \delta(z-z_0)$. The thermal energy-scaled potential of mean force, $\beta U(z)$, can be estimated from $\rho(z)$ by $\beta U(z) = -\ln[\rho(z)/\rho_{bulk}^{45°C}]$. Across systems with various intermembrane separations, the result of $\eta_D^2 \phi_D(t)$ calculated using our theory closely matches the long-time profile of $C_D(t)$ directly extracted from the MD simulation results (Figure 4a). In addition, the NGP time profiles calculated from eq 4b are also in quantitative agreement with the MD simulation results (Figure 4b). Furthermore, our theory predicts the lateral displacement distribution with a unimodal peak and a non-Gaussian tail for nanoconfined water molecules. We find the prediction of our theory in excellent agreement with the MD simulation results for the time-dependent lateral displacement distribution of water molecules at various times and separations between the membranes (Figure 4c). The agreement between theory and simulation again demonstrates the validity of our assumption underlying eq 1, that is, the most important variable that affects the lateral thermal motion of an interfacial water molecule is the distance, *z*, between a water molecule and the center of the membrane.

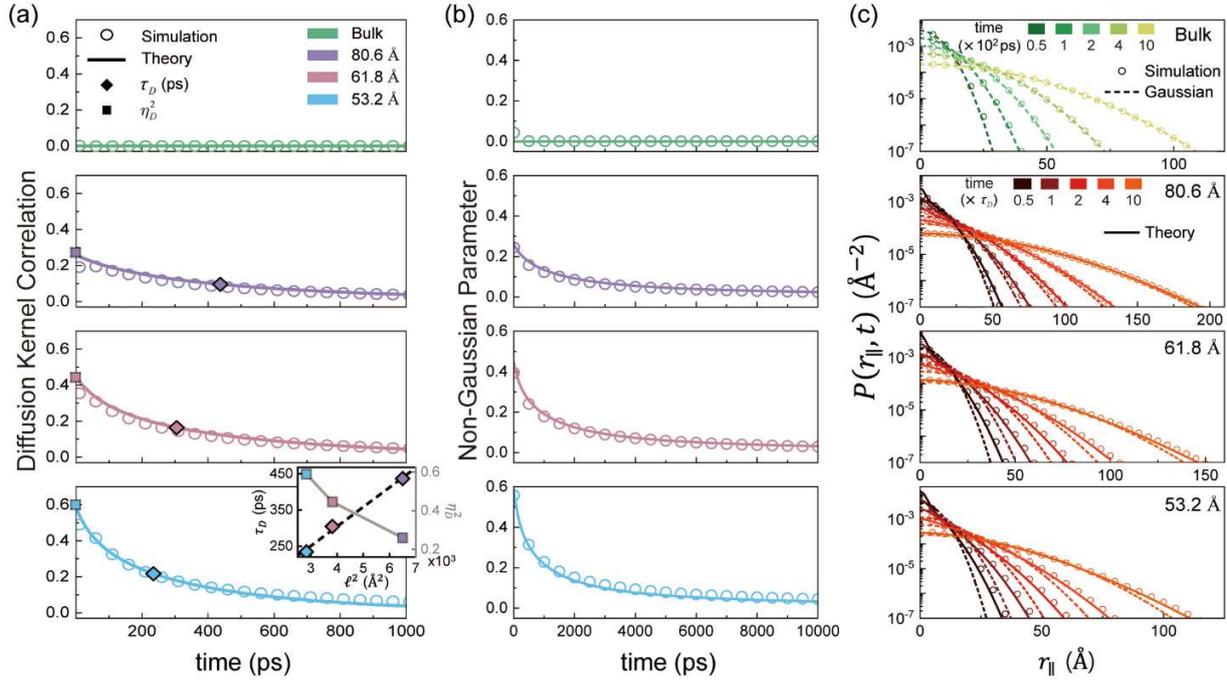

**Figure 4.** Diffusion Kernel Correlation, Non-Gaussian Parameter, and Lateral Displacement Distribution of water molecules confined between two lipid membranes with various intermembrane separations. (a) Comparison between the diffusion kernel correlation and the mean-scaled time correlation function of the lateral diffusion coefficient fluctuation: (circles) Diffusion Kernel Correlation, $C_\mathcal{D}(t)$, from the MD simulation results; (solid lines) mean-scaled TCFs, $\eta_D^2 \phi_D(t)$, of the lateral diffusion coefficient fluctuation calculated by eq 5; (diamond) relaxation time $\tau_D$ defined by $\phi_D(\tau_D) = e^{-1}$; (square) the value of the relative variance, $\eta_D^2$, of the diffusion coefficient. (Inset in (a)) Dependence of $\tau_D$ and $\eta_D^2$ on the square, $\ell^2$, of intermembrane separation. (b) Non-Gaussian parameter $\alpha_2(t)$: (circles) simulation; (solid line) eq 4b. (c) Lateral

displacement distributions at various times: (circles) simulation results; (solid lines) theoretical predictions; (dotted line) Gaussian with zero mean and variance given by $4\langle D_\parallel \rangle t$.

The relaxation time $\tau_D$ of $\phi_D(t)$, defined by $\phi_D(\tau_D) = e^{-1}$, quadratically increases with the intermembrane separation $\ell$ (see inset of Figure 4a). This follows because, for the complete relaxation of the diffusion coefficient fluctuation, water molecules must travel the entire intermembrane space, and the time taken for this process should be proportional to the mean first passage time, which quadratically increases with $\ell$ (Movie S1). On the other hand, $\eta_D^2$ decreases with $\ell$. This is because, as $\ell$ increases, the proportion of the trapped interfacial water molecules decreases, which leads to an increase in the mean lateral diffusion coefficient and a decrease in the variance of the diffusion coefficient.

We note here that, at times shorter than 0.1 ps, the NGP increases with time, which cannot be explained by eq 4b. At such short times, the DKC can be approximated by $C_\mathcal{D}(t) \cong 2\left[\langle \mathcal{D}_\parallel(t) \rangle / \langle \mathcal{D}_\parallel(0) \rangle\right]^2 = 2\phi_v^2(t)$ (see the Figure S3a in the Supporting Information), where $\phi_v(t)$ denotes the normalized VAF of water molecules. Using this approximation in eq 2, we obtained the following short-time asymptotic expression of the NGP as

$$\alpha_2(t) = \frac{1}{18}(\gamma^2 - \mu)t^2 + \mathcal{O}(t^3),  \qquad (6)$$

where $\gamma$ and $\mu$ denote $-\lim_{t\to 0} \partial \phi_v(t)/\partial t$ and $\lim_{t\to 0} \partial^2 \phi_v(t)/\partial t^2$, respectively. Explicit expressions of $\gamma$ and $\mu$ are available for our model (see the paragraph below eq S24 in the Supporting

Information). According to this result, the NGP quadratically increases with time, which quantitatively explains our simulation results for the short-time asymptotic behavior of the NGP.

The major contributor to the short-time dynamics of the NGP is the relaxation dynamics of the velocity fluctuation, and the two-time velocity auto-correlation function completely determines the NGP time profile. After the onset of Fickian diffusion, however, the relaxation dynamics of the diffusion coefficient fluctuation additionally contributes to the NGP time profile. For bulk water system, where the diffusion coefficient fluctuation is negligible, the short-time asymptotic expression of the NGP that only accounts for the velocity relaxation provides a quantitative explanation of the long-time relaxation of the NGP as well as the short-time relaxation (see the Figure S3b in the Supporting Information).

We present a physical model and transport equation that quantitatively explain the stochastic thermal motion of water molecules in intermembrane space. The lateral displacement distribution of the water molecules nanoconfined between two membranes strongly deviates from Gaussian, which originates from dynamic fluctuation in the lateral diffusion coefficient. This fluctuation occurs because the lateral diffusion coefficient of a water molecule primarily depends on its distance from the membrane center, and this distance fluctuates over time, owing to thermal motion of the water molecule in the longitudinal direction. In addition, using the molecular dynamics (MD) simulation, we investigate the dependence of the mass density, the orientation, and the lateral and longitudinal diffusion coefficients of a water molecule nanoconfined between two phospholipid membranes on its distance from the phospholipid membrane center. This study shows the presence of interfacial region within 26.6 Å from the membrane center. Water molecules in the interfacial region have a structure and dynamics far different from bulk water molecules. The properties of interfacial water molecules are robust with respect to changes in intermembrane separation. Our theory provides a unified, quantitative explanation of our MD simulation results for the mean square displacement, non-Gaussian parameter, and displacement distribution of nanoconfined water molecules. Our model is applicable or can be extended to quantitative investigation into the dynamics of transport and transport-coupled processes occurring in various nanoconfined environments.

ASSOCIATED CONTENT

**Supporting Information**

Movie S1: Heatmap of discrete Green's function (MP4)

Additional figures, tables, computational details for MD simulation and captions for Movie S1 (PDF)

(link: https://doi.org/10.1021/acs.jpclett.4c00323)

ACKNOWLEDGMENT

This work was supported by the Institute for Basic Science (IBS-R023-D1), the Creative Research Initiative Project program (RS-2015-NR011925) funded by the National Research Foundation (NRF) of Korea, and the NRF grants (NRF-2020R1A2C1102788 and RS-2023-00245431) funded by Korea government (MSIT).

**Supporting Information for**

# Transport Dynamics of Water Molecules Confined between Lipid Membranes

Minho Lee[1,2], Euihyun Lee[3], Ji-Hyun Kim[1,2], Hyonseok Hwang[6], Minhaeng Cho[4,5*], and Jaeyoung Sung[1,2*]

[1] *Creative Research Initiative Center for Chemical Dynamics in Living Cells, Chung-Ang University, Seoul 06974, Republic of Korea*

[2] *Department of Chemistry, Chung-Ang University, Seoul 06974, Republic of Korea*

[3] *Department of Chemistry, The University of Texas at Austin, TX 78757, USA*

[4] *Center for Molecular Spectroscopy and Dynamics, Institute for Basic Science (IBS), Seoul 02841, Republic of Korea*

[5] *Department of Chemistry, Korea University, Seoul 02841, Republic of Korea*

[6] *Department of Chemistry, Institute for Molecular Science and Fusion Technology, Kangwon National University, Chuncheon, Gangwon-do 24341, Republic of Korea*

Corresponding authors: jaeyoung@cau.ac.kr (J.S.), mcho@korea.ac.kr (M.C.)

# Table of contents

# Supplementary Text S1: Computational details of molecular dynamics simulations

Molecular Dynamics (MD) simulations were carried out using the AMBER 21 program package. We constructed simulation systems composed of 128 DMPC (1,2-dimyristoyl-sn-glycero-3-phosphocholine) lipids and a various number, 2560, 3840, and 6400 of water molecules, using the CHARMM-GUI membrane builder tool[1–6]. The force field parameters of DMPC and water molecules were replaced with the AMBER lipid 14[7] and SPC/E, respectively. Similarly, an MD simulation system for pure water, which is composed of 5,000 water molecules, was constructed using CHARMM-GUI solution builder[1,2,6], and the MD simulation was carried out using the same SPC/E force field. Periodic boundary conditions were applied to both the pure water and the lipid bilayer systems. In the case of the pure water system, our pure water model system with the periodic boundary conditions serves as a realistic and reliable representation of bulk water without edge effects. For the lipid bilayer system, our periodic boundary conditions effectively confine the water molecules within the lipid membranes. The particle mesh Ewald method[8] was used to calculate long-range electrostatic interactions, and the cutoff distance of 10 Å was used to calculate the Lennard-Jones interaction and the real space part of the Ewald sum. A time step of our MD simulations was set to 1 fs.

Before obtaining the simulation trajectories of the $NVT$ ensemble of our systems, we conducted an equilibration procedure. For the pure water simulation, the initial configuration of water molecules was first stabilized through energy minimization for 5,000 steps, employing both the steepest descent method and the conjugate gradient method. This was followed by a 2 ns constant $NpT$ equilibration at 1.0 atm and 318 K, using isotropic position scaling with a relaxation time of 2 ps. For temperature control, the Langevin thermostat, with a collision frequency of 1.0 ps$^{-1}$, was utilized. Subsequently, a 2 ns constant $NVT$ simulation at 318 K was carried out, using the Langevin thermostat to ensure that the system relaxed to the thermal equilibrium state. Regarding the DMPC lipid bilayer system simulation, the initial configuration of the system underwent a 10,000-step energy minimization using both the steepest descent method and the conjugate gradient method. Then, the system was rapidly heated from 0 K to 100 K over 40 ps, using the Langevin thermostat with a collision frequency of 1.0 ps$^{-1}$ and weak restraints on the lipid molecules with a force constant of 10. kcal mol$^{-1}$ Å$^{-2}$. Subsequently, a gradual heating from 100 K to 318 K over 2 ns was performed with the same thermostat setting and restraints. After the heating process, a 50 ns *NpT* simulation was carried out at 318 K, using the Langevin thermostat with anisotropic pressure scaling (1 atm) without restraining lipids. From the results of the last process, the values of the separation, $\ell$, between two lipid membranes, i.e., system box length along the perpendicular direction to the membrane surface, were determined. The values of the intermembrane separations, $\ell$, are 53.2 Å, 61.8 Å, and 80.6 Å for our DMPC lipids simulation systems containing 2560, 3840, and 6400 water molecules, respectively.

After the equilibration procedure, production runs were carried out for our systems at 318 K under constant *NVT* conditions with the following procedures:

(1) To calculate the mean square displacement (MSD) and non-Gaussian parameter (NGP) for the lateral displacement of water molecules, shown in Figure 2, we conducted the MD simulation for each system using two different recording time intervals. The simulation trajectories were recorded every 10 fs for the first 1 ns-long *NVT* simulations to investigate the short-time dynamics of water molecules. Afterward, the trajectories were saved every 10 ps over the following 99 ns-long *NVT* simulation to investigate the long-time dynamics of the system. This 1 ns-long NVT simulations were repeated seven times for each intermembrane separation of our system.

76    (2) To obtain the z-dependent profile of the lateral diffusion coefficient, $D_{\parallel}(z)$, shown in
77        Figure S1, we conducted additional simulation runs using umbrella sampling. The
78        simulation trajectories were recorded at every 10 ps during the 1 μs-long *NVT*
79        simulations. A more detailed description of this simulation process is presented in the
80        caption of Figure S1.

81    (3) To obtain the z-dependent profile, the longitudinal diffusion coefficient, $D_{\perp}(z)$,
82        profiles shown in Figure S2, using the method presented in Ref.[9–11], we conducted
83        another set of simulations where the simulation trajectories were saved at intervals of
84        1 ps for the 1 μs-long *NVT* simulations. A more detailed description of this simulation
85        is presented in the caption of Figure S2.

# Supplementary Text S2: Derivation of analytic expressions for the MSD and NGP

Let us first obtain the analytic expression for the second and fourth moments of the lateral displacement of water molecules in the intermembrane space, starting from eq 1. On the left-hand-side of eq 1, $\dot{\hat{p}}(\mathbf{r}_{\|},z,s)$ can be replaced by $\dot{\hat{p}}(\mathbf{r}_{\|},z,s) = s\hat{p}(\mathbf{r}_{\|},z,s) - p(\mathbf{r}_{\|},z,0)$. Here, $p(\mathbf{r}_{\|},z,0)$ denotes the initial condition of the joint probability density, given by $p(\mathbf{r}_{\|},z,0) = \delta(\mathbf{r}_{\|} - \mathbf{r}_{\|,0})P_{eq}(z)$, where, $\mathbf{r}_{\|,0}$ and $P_{eq}(z)$ denote the initial lateral position vector and the equilibrium distribution of water molecule along the $z$-axis, respectively. When $\mathbf{r}_{\|,0}$ is chosen to be the origin of our coordinate, i.e., $\mathbf{r}_{\|,0} = \mathbf{0}$, $\mathbf{r}_{\|}$ represents the displacement vector. By taking the Fourier transform of eq 1, we obtain

$$\dot{\hat{\tilde{p}}}(\mathbf{k}_{\|},z,s) = s\hat{\tilde{p}}(\mathbf{k}_{\|},z,s) - P_{eq}(z) = -\hat{\mathcal{D}}_{\|}(z,s)k_{\|}^2\hat{\tilde{p}}(\mathbf{k}_{\|},z,s) + L(z)\hat{\tilde{p}}(\mathbf{k}_{\|},z,s). \qquad (S1)$$

Here, $\hat{\tilde{p}}(\mathbf{k}_{\|},z,s)$ and $k_{\|}$ represent the Fourier transformation of $\hat{p}(\mathbf{r}_{\|},z,s)$, defined by $\hat{\tilde{p}}(\mathbf{k}_{\|},z,s) = \int_0^\infty d\mathbf{r}_{\|} e^{i\mathbf{k}_{\|}\cdot\mathbf{r}_{\|}}\hat{p}(\mathbf{r}_{\|},z,s)$, and the magnitude of the wave vector, $\mathbf{k}_{\|}$, i.e. $k_{\|} = |\mathbf{k}_{\|}|$, respectively.

The first two non-vanishing moments, $\Delta_2(t)[\equiv \langle\Delta\mathbf{r}_{\|}(t)^2\rangle]$ and $\Delta_4(t)[\equiv \langle\Delta\mathbf{r}_{\|}(t)^4\rangle]$, of the distribution of the lateral displacement $\Delta\mathbf{r}_{\|}(t)[\equiv \mathbf{r}_{\|}(t) - \mathbf{r}_{\|}(0)]$ are related to z-dependent displacement distribution, $p(\mathbf{r}_{\|},z,t)$, by

$$\Delta_n(t) = \int_0^\ell dz \int d\mathbf{r}_{\|}(r_{\|})^n p(\mathbf{r}_{\|},z,t) = \int_0^\ell dz\, \Delta_n(z,t), \qquad (S2)$$

104    where $\Delta_n(z,t)$ is defined by $\Delta_n(z,t) = \int d\mathbf{r}_\| (r_\|)^n p(\mathbf{r}_\|,z,t) = 2\pi \int_0^\infty dr_\| (r_\|)^{n+1} p(\mathbf{r}_\|,z,t)$. The

105    expression of $\Delta_n(z,t)$ can be obtained by the second and fourth derivatives of $\tilde{p}(\mathbf{k}_\|,z,t)$

106    with respect to $k_\|$ and setting $k_\| = 0$ in the resulting equation, that is,

$$\partial_{k_\|}^q \tilde{p}(k_\|,z,t)\Big|_{k_\|=0} = \partial_{k_\|}^q \int d\mathbf{r}_\| e^{ik_\| r_\| \cos\theta} \tilde{p}(\mathbf{r}_\|,z,t)\Big|_{k_\|=0}$$
$$= i^q \int d\mathbf{r}_\| \left[(r_\| \cos\theta)^q \tilde{p}(\mathbf{r}_\|,z,t)\right]$$
$$= i^q \int_0^\infty dr_\| \int_0^{2\pi} d\theta\, r_\| \left[(r_\| \cos\theta)^q \tilde{p}(\mathbf{r}_\|,z,t)\right]$$
$$= \begin{cases} -\Delta_2(z,t)/2, & \text{for } q=2 \\ 3\Delta_4(z,t)/8, & \text{for } q=4 \end{cases}$$

(S3)

108    where $\theta$ denotes the angle between the two vectors, $\mathbf{k}_\|$ and $\mathbf{r}_\|$, defined by

109    $\cos\theta = \mathbf{k}_\| \cdot \mathbf{r}_\| / (k_\| r_\|)$. Taking the mathematical operation, $\partial_{k_\|}^2 (...)_{k_\|=0}$, on both sides of eq S1

110    and using eq S3, we obtain

$$\hat{\Delta}_2(z,s) = 4(s - L(z))^{-1} \hat{\mathcal{D}}_\|(z,s) P_{eq}(z)/s.$$  (S4)

112    Equation S4 can be written as

$$\hat{\Delta}_2(z,s) = 4\int_0^\ell dz_0 \hat{G}(z,s|z_0)\hat{\mathcal{D}}_\|(z_0,s) P_{eq}(z_0)/s.$$  (S5)

114    where $\hat{G}(z,s|z_0)\left[=(s-L(z))^{-1}\delta(z-z_0)\right]$ denotes the Laplace transform of Green's function

115    $G(z,t|z_0)$ defined by $\partial_t G(z,t|z_0) = L(z) G(z,t|z_0)$, with the initial condition,

116    $G(z,0|z_0) = \delta(z-z_0)$. The Green's function represents the conditional probability that a water

117    molecule initially located at $z_0$ is found at $z$ at time $t$. Integrating both sides of eq S5 over $z$,

118    and using the normalization condition, $\int_0^\ell dz\, G(z,t|z_0) = 1$, we obtain

119 $$\hat{\Delta}_2(s) = 4\langle\hat{\mathcal{D}}_\|(s)\rangle/s^2. \quad (S6)$$

120   The analytic expression of $\Delta_4(t)$ can be obtained by following a similar line of

121 derivation. Taking the mathematical operation, $\partial_{k_\|}^4(...)_{k_\|=0}$, on both sides of eq S1 and using

122 eq S3, we obtain

123 $$\hat{\Delta}_4(z,s) = 64\int_0^\ell dz_1 \int_0^\ell dz_0 \hat{G}(z,s|z_1)\hat{\mathcal{D}}_\|(z_1,s)\hat{G}(z_1,s|z_0)\hat{\mathcal{D}}_\|(z_0,s)P_{eq}(z_0)/s. \quad (S7)$$

124 Integrating both sides of eq S7 over $z$, and using the normalization condition,

125 $\int_0^\ell dz\, G(z,t|z_0) = 1$, we obtain

126 $$\hat{\Delta}_4(s) = \frac{64}{s^3}\langle\hat{\mathcal{D}}_\|(s)\rangle^2\left(1 + s\hat{C}_\mathcal{D}(s)\right) = 4s\hat{\Delta}_2(s)^2\left(1 + s\hat{C}_\mathcal{D}(s)\right), \quad (S8)$$

127 where $C_\mathcal{D}(t)$ denotes the lateral diffusion kernel correlation (DKC) defined by

128 $$\hat{C}_\mathcal{D}(s) = \int_0^\ell dz\int_0^\ell dz_0 \frac{\delta\hat{\mathcal{D}}_\|(z,s)}{\langle\hat{\mathcal{D}}_\|(s)\rangle}\hat{G}(z,s|z_0)\frac{\delta\hat{\mathcal{D}}_\|(z_0,s)}{\langle\hat{\mathcal{D}}_\|(s)\rangle}P_{eq}(z_0). \quad (S9)$$

129 Here, the lower and upper bounds, 0 and $\ell$, of the integral denote the positions of two different

130 membrane centers. That is to say, $\ell$ denotes the separation between the two lipid membrane

131 centers (see Figure 1).

132   After the onset of Fickian diffusion, the MSD linearly increases with time because the

133 mean diffusion kernel, $\langle\mathcal{D}_\|(z,t)\rangle$, is negligibly small at the time scale of Fickian diffusion,

134 where we have $\Delta_2(t) = 4\int_0^t d\tau(t-\tau)\langle\mathcal{D}_\|(\tau)\rangle \cong 4t\int_0^\infty \langle\mathcal{D}_\|(\tau)\rangle$. After the onset of Fickian

135 diffusion, the diffusion kernel, $\hat{\mathcal{D}}_\|(z,s)$, can be replaced by its small-s limit value, $\hat{\mathcal{D}}_\|(z,0)$,

136 which is nothing but the lateral diffusion coefficient, $D_\|(z)$, of water molecules at longitudinal

137 position $z$. Additionally, after the onset of Fickian diffusion, $L(z)$ in eq 1 can be
138 approximated by the Smoluchowski operator, defined as, $L_{SM}(z) = \partial_z[D_\perp(z)(\partial_z + \partial_z \beta U(z))]$.
139 Here, $D_\perp(z)$ and $\beta U(z)$ respectively denote the $z$-dependent diffusion coefficient
140 associated with the thermal motion of water molecules in the longitudinal direction and the
141 thermal energy-scaled potential of mean force. Throughout $\beta$ denotes $\beta = 1/k_B T$ where
142 $k_B$ and $T$ denote the Boltzmann constant and temperature, respectively. Therefore, at the time
143 scales of Fickian diffusion, the exact expression of the first two non-vanishing moments, eqs
144 S6 and S8, can be approximated as

$$\Delta_2(t) = 4\langle D_\parallel \rangle t, \quad (S10)$$

$$\Delta_4(t) = 32\langle D_\parallel \rangle^2 \left[ t^2 + 2\eta_D^2 \int_0^t dt'(t-t')\phi_D(t') \right]. \quad (S11)$$

147 Here, $\langle D_\parallel \rangle$ and $\eta_D^2 [= \langle \delta D_\parallel(z)^2 \rangle / \langle D_\parallel \rangle^2]$ respectively denote, the mean diffusion coefficient
148 and the relative variance of the $z$-dependent lateral diffusion coefficient. $\phi_D(t)$ represents the
149 normalized time-correlation function (TCF) of the lateral diffusion coefficient fluctuation given
150 by

$$\phi_D(t) = \frac{\langle \delta D_\parallel(t) \delta D_\parallel(0) \rangle}{\langle \delta D_\parallel^2 \rangle} = \langle \delta D_\parallel^2 \rangle^{-1} \int_0^\ell dz \int_0^\ell dz_0 \delta D_\parallel(z) G_{SM}(z,t|z_0) \delta D_\parallel(z_0) P_{eq}(z_0), \quad (S12)$$

152 where $G_{SM}(z,t|z_0)$ designates Green's function defined by $\partial_t G_{SM}(z,t|z_0)$
153 $= L_{SM}(z) G_{SM}(z,t|z_0)$, with the initial condition, $G_{SM}(z,0|z_0) = \delta(z-z_0)$.

154 Substituting the eqs S10 and S11 into the definition of the NGP, $\alpha_2(t)$
155 $[\equiv \Delta_4(t)/(2\Delta_2(t)^2) - 1]$, we obtain the analytical expressions of the NGP time profile as

156 $$\alpha_2(t) = \frac{2\eta_D^2}{t^2} \int_0^t dt'(t-t')\phi_D(t'). \tag{S13}$$

157 Equations S6, S8, S10, and S13 are equivalent to eqs 2a, 2b, 4a, and 4b in the main text,
158 respectively.

Figure S1: z-dependent profiles of lateral diffusion coefficient

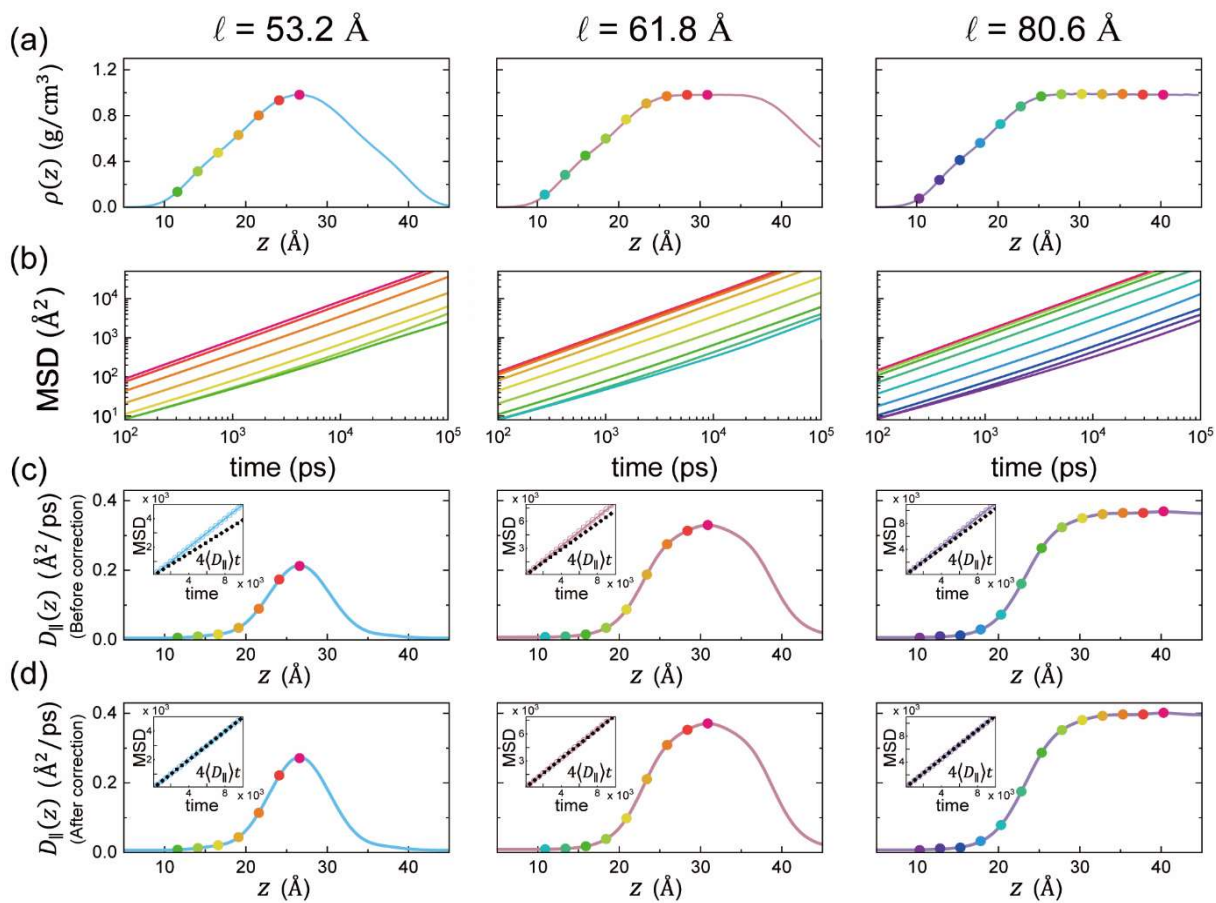

To determine the z-dependent profile of the lateral diffusion coefficient, we obtain the $z$-dependence of the MSD of water molecules in the intermembrane space. For this purpose, we estimate the MSD time profile of water molecules within each layer region defined by $z_0 - 1.5\text{Å} \leq z \leq z_0 + 1.5\text{Å}$, systematically changing the center position $z_0$ of the layer. It is difficult to estimate the long-time behavior of the MSD accurately, because very few water molecules remain in the initial layer at long times[11]. To circumvent this difficulty, we employ a constrained MD simulation in which we randomly choose approximately 10 % of water molecules within a layer and apply Harmonic potential, $U(z) = k(z-z_0)^2$ to the chosen water molecules. Here, $k$ represents the spring constant whose value is set to be 1.25 [Kcal/(mol

170    ·Å$^2$)].

171    For the first round of our constrained simulation, we set the value of $z_0$ to be $\ell/2$. Then
172    we repeatedly perform the constrained simulation, systematically changing the value of $z_0$ by
173    2.5 Å, until we span the entire intermembrane space. The position of $z_0$ for each simulation
174    is represented by a colored dot in Figure S1a. In the constrained MD simulation, the MD
175    trajectories were recorded every 10 ps during 1μs-long *NVT* simulations. The number of water
176    molecules constrained by the harmonic potential is about 6% for each system, which amounts
177    to 403, 240, and 168 for the system with $\ell$ = 80.6 Å, 61.8 Å, and 53.2 Å.

178    From the trajectories of water molecules constrained by the harmonic potential, we
179    obtained the MSDs of the lateral water displacement for every layer of the simulation system
180    with different intermembrane separations (Figure S1b). Each solid line in Figure S1b represents
181    the MSD obtained from the constrained MD simulation with $z_0$ at the position marked by the
182    dot of the same color in Figure S1a. From the long-time MSD profile, we estimate the lateral
183    diffusion coefficient at each *z* position by $D_\parallel(z) = \lim_{t \to \infty} \Delta_2(z,t)/4t$. These values are shown
184    as circle symbols in Figure S1c. The resulting profile of the lateral diffusion coefficient can
185    well be fitted to the muti-Gaussian function,

$$D_\parallel(z) = y_0 + \sum_{i=1}^{n} \frac{a_i}{\omega_i \sqrt{\pi/2}} \exp\left(-2\left(\frac{z - z_{c,i}}{\omega_1}\right)^2\right), \quad \text{(S14)}$$

187    as shown as solid lines in Figure S1c. The optimized parameters for these fittings are provided
188    in Table S1. Using the multi-Gaussian representation of $D_\parallel(z)$, we calculate the mean lateral
189    diffusion coefficient, $\langle D_\parallel \rangle$, by

190   $$\langle D_\parallel \rangle = \int_0^\ell dz D_\parallel(z) P_{eq}(z) \qquad (S15)$$

191   with $P_{eq}(z) = \rho(z) \Big/ \int_0^\ell dz \rho(z)$. Here, $\rho(z)$ denotes the mass density profile of water

192   molecules, shown in Figure S1a.

193   To test the accuracy of $D_\parallel(z)$ obtained from our constrained MD simulations, we

194   compared $4\langle D_\parallel \rangle t$ with the MSD obtained from the simulation of the entire system as shown

195   in insets of Figure S1c. These results clearly show that the estimation of $D_\parallel(z)$ and $\langle D_\parallel \rangle$

196   from our constrained MD simulation is not perfectly accurate. $\langle D_\parallel \rangle$ estimated from our

197   constrained MD simulation is slightly smaller than the true value of $\langle D_\parallel \rangle$ estimated from the

198   MSD of the entire system. This discrepancy is expected because 10% of the water molecules

199   are constrained in each layer by the fictitious harmonic potential, and their transport dynamics

200   would not be exactly the same as the transport dynamics of the free water molecules moving

201   across various layers.

202   We resolve this issue by introducing a correction factor, $c$, to $D_\parallel(z)$ estimated from our

203   constrained MD simulation in such a way that the mean lateral diffusion coefficient calculated

204   by $c \int_0^\ell dz D_\parallel(z) P_{eq}(z)$ is the same as the true value of $\langle D_\parallel \rangle$ estimated from the MSD of the

205   entire system (see insets of Figure S1d). The values of the correction factor are 1.087, 1.124,

206   and 1.276 for the systems with $\ell = 80.6$ Å, 61.8 Å, and 53.2 Å, respectively. The value of the

207   corrected lateral diffusion coefficient, $cD_\parallel(z)$, are represented by solid lines in Figure S1d and

208   Figure 3c in the main text.

209   We use the corrected lateral diffusion coefficient profiles in calculating the theoretical

210   results depicted in Figure 4 of the main text. The agreement between our theoretical results and the MD simulation results for the time profiles of the NGP, the TCF of the lateral diffusion coefficient, and the lateral displacement distribution at various times confirms the accuracy of our corrected lateral diffusion coefficient profiles.

Figure S2: z-dependent profiles of longitudinal diffusion coefficient

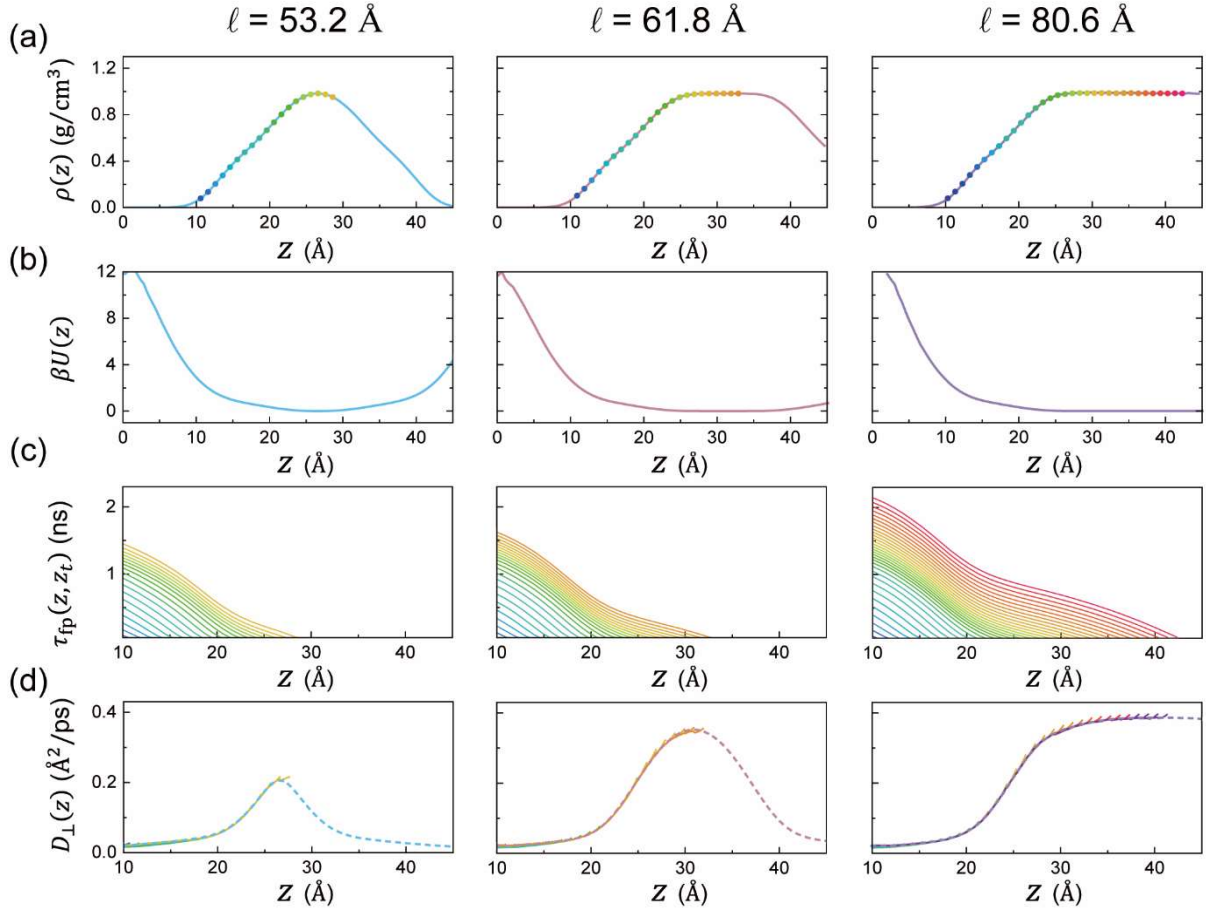

We obtain the longitudinal diffusion coefficient profile, $D_\perp(z)$, from the mean-first passage time profile (MFPT), using the method developed in Ref.[9–11]. The MFPT, $\tau_{\text{fp}}(z, z_t)$, denotes the average time required for a water molecule initially located at $z$ to arrive at a target position $z_t$. This method is based on the analytic results of the previous theories that represent $\tau_{\text{fp}}(z, z_t)$ as a functional of $D_\perp(z)$ and the potential of mean force, $\beta U(z)$[12,13]. According to Ref. 9, $D_\perp(z)$ can be obtained from

$$D_\perp(z) = -\frac{e^{\beta U(z)}}{\partial \tau_{\text{fp}}(z, z_t)/\partial z} \int_{z_{\text{refl}}}^{z} dz' \, e^{-\beta U(z')}. \quad (S16)$$

Here, thermal energy scaled potential of mean force, $\beta U(z)$, of water molecules can be calculated from the density profile of water molecules, shown in Figure S2a, i.e., $\beta U(z) \equiv -\log[\rho(z)/\rho_{bulk}^{45°C}]$ with $\rho_{bulk}^{45°C}$ being the density of bulk water at 45°C. In eq S16, $z_{refl}$ designates the position of the reflecting boundary. As the number of water molecules passing through the lipid bilayer membrane is negligibly small throughout our MD simulation, we set the reflecting boundary at the membrane center, i.e., $z_{refl} = 0$.

To calculate $D_\perp(z)$ using eq S16, we computed the z-dependent profile of $\tau_{fp}(z, z_t)$ using our MD trajectories, systematically changing the value of $z_t$ by 1 Å. The various positions of $z_t$ for each system are represented by the dots of different colors in Figure S2a. The z-dependent profiles of MFPT for various values of $z_t$ have the same slope through the entire z range, as shown in Figure S2c. Each solid line in Figure S2c represents $\tau_{fp}(z, z_t)$ with the target position, $z_t$, represented by the dot of the same color in Figure S2a. As $\partial_z \tau_{fp}(z, z_t)$ is independent of $z_t$, so is $D_\perp(z)$ calculated from eq S15 (see Figure S2d). The profile of $D_\perp(z)$ was well fitted to a multi-Gaussian function,

$$D_\perp(z) = y_0 + \sum_{i=1}^{n} \frac{a_i}{\omega_i \sqrt{\pi/2}} \exp\left(-2\left(\frac{z - z_{c,i}}{\omega_i}\right)^2\right), \quad (S17)$$

shown as dotted lines in Figure S2d. The optimized parameters for these fittings are provided in Table S2. These best-fitted results, shown as dotted lines in Figure S2d, are represented by the solid lines in Figure 3d of the main text.

S16

## Figure S3: Diffusion Kernel Correlation and Non-Gaussian Parameter

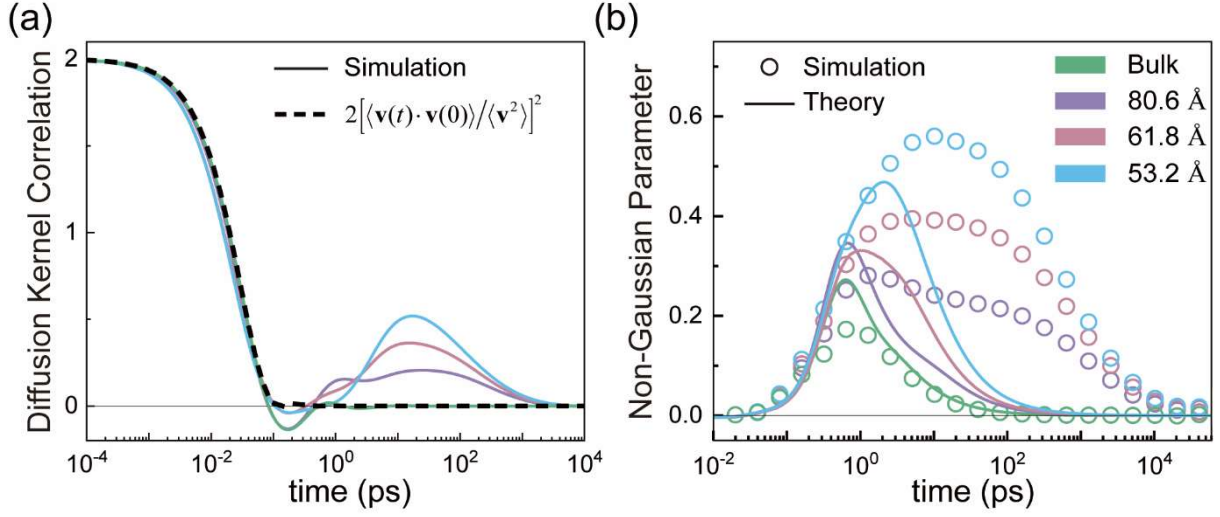

The lateral diffusion kernel correlation (DKC), $C_\mathcal{D}(t)$, is a key dynamic quantity that characterizes the environment-coupled fluctuation of water molecule's motility in the lateral direction. Based on the eq S8, we express the DKC in the Laplace domain as

$$\hat{C}_\mathcal{D}(s) = \frac{\hat{\Delta}_4(s)}{4s^2\hat{\Delta}_2(s)^2} - \frac{1}{s} \tag{S18}$$

To extract the time-profile of the DKC of water molecules by performing numerical inverse Laplace transform of eq S18, it is convenient to have analytic expressions of the first two non-vanishing moments, $\Delta_2(t)$ and $\Delta_4(t)$, of the distribution of the lateral displacement of water molecules.

The time profile of the MSD, $\Delta_2(t)$, is well represented by the following formula according to Ref.14:

$$\Delta_2(t) = 4\frac{k_BT}{M\gamma_0^2}c_0\left(\gamma_0 t - 1 + e^{-\gamma_0 t}\right) + 4\frac{k_BT}{M}\sum_{i=1}^{2}\frac{c_i}{\omega_{0,i}^2}\left[1 - e^{-\gamma_i t}\left(\cosh\omega_i t + \frac{\gamma_i}{\omega_i}\sinh\omega_i t\right)\right] \tag{S19}$$

This formula represents the MSD of a bead in a Gaussian polymer composed of three beads in

S17

a high friction regime. The first and second terms on the R.H.S of eq 19 account for the contribution from the unbound mode and bound modes, respectively. $c_i$ designates the weight coefficient of *i*th mode, which satisfies the following condition $\sum_{j=0}^{2} c_j = 1$. The optimized parameters for this fitting are provided in Table S3.

The analytic expression of the fourth moment, $\Delta_4(t)$, is obtained from

$$\Delta_4(t) = 2\Delta_2(t)^2 [1 + \alpha_2(t)] \tag{S20}$$

which is obtained from the definition of the NGP. The simulation result for the time profile of the NGP, or $\alpha_2(t)$, could be well fitted to the following function (see Figure 2):

$$\alpha_2(t) \cong \sum_{i=1}^{2\,\text{or}\,3} a_i \exp\left[-\left(\log_{10} t - b_i\right)^2 / c_i\right]. \tag{S21}$$

The optimized parameters for this fitting are provided in Table S4. Substituting eqs S19 and S21 into eq S20, we obtain the fully analytic expression of the fourth moment.

Taking the Laplace transforms of $\Delta_2(t)$ and $\Delta_4(t)$, and substituting the results into eq S18, we obtain the expression of DKC in the Laplace domain. To obtain the value of DKC at a specific timepoint, *t*, we perform the numerical Laplace inversion of eq S18, by using the Stehfest algorithm. These results are shown in Figure S3 and Figure 4a. The DKC values in various conditions are shown as solid lines in Figure S3a. At short times, where MSD is not linear in time yet, the DKC can be approximated by $C_{\mathcal{D}}^{(short)}(t) = 2\left[\langle \mathbf{v}(t) \cdot \mathbf{v}(0) \rangle / \langle \mathbf{v}^2 \rangle\right]^2 = 2\phi_{\mathbf{v}}^2(t)$ [14]. Here, **v** denotes the water molecule's velocity vector in the lateral direction, and the velocity autocorrelation function, $\langle \mathbf{v}(t) \cdot \mathbf{v}(0) \rangle$, is equivalent to twice the lateral diffusion kernel, $\langle \mathcal{D}_\parallel(t) \rangle$ [14]. From eqs S6 and S19, an analytic expression of the lateral diffusion kernel

276 can be obtained as

$$\frac{\langle \mathcal{D}_\parallel(t) \rangle}{\langle \mathcal{D}_\parallel(0) \rangle} \left( = \frac{\langle \mathbf{v}(t) \cdot \mathbf{v}(0) \rangle}{\langle |\mathbf{v}|^2 \rangle} \right) = c_0 e^{-\gamma_0 t} + \sum_{i=1}^{2} c_i e^{-\gamma_i t} \left[ \cosh \omega_i t + \frac{\gamma_i}{\omega_i} \sinh \omega_i t \right] \quad (S22)$$

278 where $\langle \mathcal{D}_\parallel(0) \rangle$ denotes $k_B T / M$.

279 To verify the correctness of $C_\mathcal{D}(t) \cong 2\phi_\mathbf{v}^2(t)$, we made a comparison between $2\phi_\mathbf{v}^2(t)$ of
280 bulk water, depicted as the dashed line in Figure S3, and the time profile of the DKCs. The
281 agreement between $2\phi_\mathbf{v}^2(t)$ and the DKCs at short times confirms the validity of our short-
282 time asymptotic expression of the DKC. As shown in Figure S3a, the short-time dynamics of
283 the DKC is largely independent of the degree of confinement, or $\ell$.

284 The short-time asymptotic expression of the DKC enables us to obtain the short-time
285 asymptotic behavior of $\alpha_2(t) \propto t^2$. Using the asymptotic expression of the DKC in the exact
286 analytic expression of the fourth moment, given in eq S8, we obtain

$$\hat{\Delta}_4(s) \cong 4s\hat{\Delta}_2(s)^2 [1 + s\hat{C}_\mathcal{D}^{(short)}(s)]. \quad (S23)$$

288 Substituting the time-domain version of eq S23 into the definition of the NGP, $\alpha_2(t)$
289 $\left[ \equiv \Delta_4(t)/(2\Delta_2(t)^2) - 1 \right]$, we obtain the following short-time asymptotic expression of $\alpha_2(t)$:

$$\alpha_2(t) = \frac{1}{18}(\gamma^2 - \mu)t^2 + \mathcal{O}(t^3) \quad (S24)$$

291 In eq S24, $\gamma$ and $\mu$ are given by $\gamma = -\lim_{t \to 0} \partial \phi_\mathbf{v}(t)/\partial t = c_0 \gamma_0 + 2\sum_{i=1}^{n} c_i \gamma_i$ and
292 $\mu = \lim_{t \to 0} \partial^2 \phi_\mathbf{v}(t)/\partial t^2 = c_0 \gamma_0^2 + \sum_{i=1}^{n} c_i (3\gamma_i^2 + \omega_i^2)$, respectively[14]. These results are shown as
293 the solid lines in Figure S3b. The short-time asymptotic behavior of the NGP given in eq S24
294 is found to be in good quantitative agreement with our MD simulations results.

Figure S4: Relationship between DKC and NGP at long times

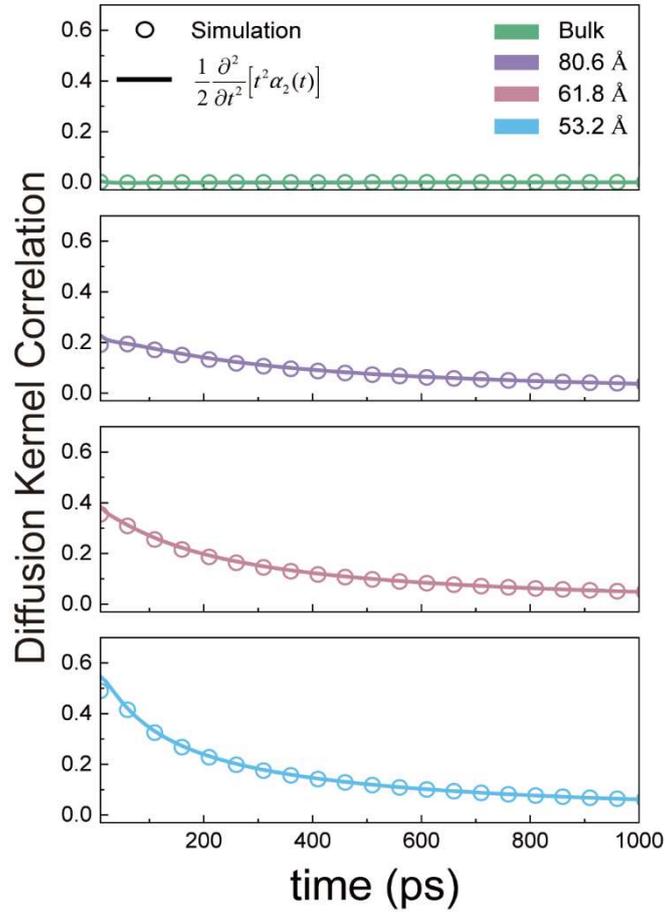

The relaxation dynamics of the DKC, $C_\mathcal{D}(t)$, determines the NGP time profile of water molecules in the intermembrane space after the onset of Fickian diffusion, where the time profile of $C_\mathcal{D}(t)$ can be obtained by the NGP time profile, i.e.,

$$C_\mathcal{D}(t) \cong \eta_D^2 \phi_D(t) = \frac{1}{2}\frac{\partial^2}{\partial t^2}[t^2 \alpha_2(t)]. \tag{S25}$$

Solid lines in Figure S4 represent the time profiles of $C_\mathcal{D}(t)$ obtained from eqs S25 and S21. Circles represent the time profiles of $C_\mathcal{D}(t)$ extracted from our MD simulation results, using the method described in the paragraph below Figure S3.

# Caption for Movie S1: Heatmap of discrete Green's function

As we mentioned in the main text, linear dependence between $\tau_D$ and $\ell^2$, as shown in the inset of Figure 4a, arises because the time taken for water molecules to traverse the entire intermembrane space increases quadratically with separation, $\ell$, between center membranes. To show the stochastic transport dynamics of water molecules at various initial positions, we present a movie showing the time-dependent changes in the probability distribution of water molecules in the intermembrane space. For this purpose, we first discretize the intermembrane space into three regions: two interfacial regions near the two membranes confining the water molecules, defined by $z < z_c$ and $z > \ell - z_c$ with $z_c = 23.6$ Å, and one bulk water-like region between the two interfacial regions (see Figure 3 in the main text). The bulk water-like region does not exist for our system with $\ell = 53.2$ Å; however, it exists for our system with a greater value of intermembrane separation $\ell$.

We identify the entire bulk water-like region as a single layer named layer 0. Each interfacial water region is divided into 5 layers. Specifically, the interfacial water region on the left side is divided into the following layers: (i) 0 Å $\leq z <$ 15.2 Å, (ii) 15.2 $\leq z <$ 18 Å, (iii) 18 Å $\leq z <$ 20.8 Å, (iv) 20.8 Å $\leq z <$ 23.6 Å, and (v) 23.6 Å $\leq z < z_c$ (= 26.6 Å). These layers are respectively labeled as -5, -4, -3, -2, and -1 layers. Similarly, the interfacial water region on the right side is divided from the membrane center on the right side with identical widths, and the corresponding layers are designated as 5, 4, 3, 2, 1 layer, respectively. Consequently, the total number of layers is 11 for $\ell =$ 80.6 Å and $\ell =$ 61.8 Å, and 10 for $\ell =$ 53.2 Å.

In Supplemental Movie, we show how the spatial distribution of water molecules in each layer changes over time. For this purpose, we present heat maps, each of which represents the probability distribution of water molecules for our system at a given time. In each heat map, the color of the cell in the $n$-th column and $m$-th row represents the probability, $g(n,t|m)$, that a water molecule initially located in the $m$-th layer is found in the $n$-th layer at time $t$. By definition, the initial condition of $g(n,t|m)$ is given by $\lim_{t \to 0} g(n,t|m) = \delta_{nm}$ where $\delta_{nm}$ designates Kronecker's delta.

The evolving heatmap patterns, as shown in Supplemental Movie, clearly demonstrate that the time required to reach the equilibrium state increases with the length of $\ell$. Specifically, in the system with $\ell = 80.6$ Å, water molecules located in the interfacial region do not transfer to the opposite side of the interfacial water region within the first 50 ps. This behavior can be attributed to the increased size of the bulk-water-like region, which hinders the fast exchange of water molecules between the interfacial water regions.

Tables S1-S4

| $\ell$ (Å) | 80.6 | 61.8 | 53.2 |
|---|---|---|---|
| $y_0$ | 0.007 | 0.008 | 0.005 |
| $a_1$ | 0.636 | 0.374 | 0.029 |
| $a_2$ | 0.636 | 0.374 | 0.029 |
| $a_3$ | 5.061 | 4.412 | 1.951 |
| $a_4$ | 5.061 | | |
| $a_5$ | 0.721 | | |
| $\omega_1$ | 5.944 | 4.797 | 4.079 |
| $\omega_2$ | 5.944 | 4.797 | 4.079 |
| $\omega_3$ | 12.226 | 11.100 | 7.449 |
| $\omega_4$ | 12.226 | | |
| $\omega_5$ | 7.060 | | |
| $z_{c,1}$ | -14.750 | -6.131 | -10.790 |
| $z_{c,2}$ | 14.750 | 6.131 | 10.790 |
| $z_{c,3}$ | -8.002 | 0 | 0 |
| $z_{c,4}$ | 8.002 | | |
| $z_{c,5}$ | 0 | | |

**Table S1.** Lateral diffusion coefficient fitting parameters (eq S14).

| $\ell$ (Å) | **80.6** | **61.8** | **53.2** |
|---|---|---|---|
| $y_0$ | 0.021 | 0.018 | 0.014 |
| $a_1$ | 0.063 | -0.400 | 0.973 |
| $a_2$ | 0.058 | -0.400 | 0.953 |
| $a_3$ | 3.122 | 5.460 | 0.055 |
| $a_4$ | 2.286 | | |
| $a_5$ | 1.015 | | |
| $a_6$ | 0.974 | | |
| $a_7$ | 2.138 | | |
| $a_8$ | 1.826 | | |
| $\omega_1$ | 5.873 | 6.282 | 5.554 |
| $\omega_2$ | 5.856 | 6.281 | 16.780 |
| $\omega_3$ | 8.568 | 12.991 | 1.901 |
| $\omega_4$ | 8.089 | | |
| $\omega_5$ | 5.958 | | |
| $\omega_6$ | 6.136 | | |
| $\omega_7$ | 7.413 | | |
| $\omega_8$ | 13.114 | | |
| $z_{c,1}$ | -22.820 | -9.125 | 0 |
| $z_{c,2}$ | 22.777 | 9.125 | 0 |
| $z_{c,3}$ | -11.448 | 0 | 0 |
| $z_{c,4}$ | 11.914 | | |
| $z_{c,5}$ | -5.603 | | |
| $z_{c,6}$ | 5.456 | | |
| $z_{c,7}$ | -0.453 | | |
| $z_{c,8}$ | 5.456 | | |

**Table S2.** Longitudinal diffusion coefficient fitting parameters (eq S17).

|  | Bulk | $\ell=80.6$ Å | $\ell=61.8$ Å | $\ell=53.2$ Å |
|---|---|---|---|---|
| $c_0$ | 0.0552 | 0.0624 | 0.0126 | 0.0970 |
| $c_1$ | 0.0386 | 0.0403 | 0.4937 | 0.4515 |
| $c_2$ | 0.9062 | 0.8973 | 0.4937 | 0.4515 |
| $\gamma_0$ | 2.0867 | 3.2698 | 0.9116 | 11.4206 |
| $\gamma_1$ | 2.1041 | 4.9715 | 9.8813 | 11.4206 |
| $\gamma_2$ | 9.1332 | 9.5602 | 11.0886 | 11.7578 |
| $\omega_{0,1}$ | 2.1041 | 1.8450 | 9.8813 | 11.4206 |
| $\omega_{0,2}$ | 9.1332 | 9.5602 | 4.4480 | 6.0659 |
| $\omega_1$ | 0.0017 | 4.6165 | 0.0001 | 0.0002 |
| $\omega_2$ | 0.0018 | 0.0026 | 10.1574 | 6.0659 |

**Table S3.** Diffusion Kernel fitting parameters (eq S19).

|  | Bulk | $\ell=80.6$ Å | $\ell=61.8$ Å | $\ell=53.2$ Å |
|---|---|---|---|---|
| $a_1$ | 0.166 | 0.247 | 0.269 | 0.263 |
| $a_2$ | 0.021 | 0.105 | 0.131 | 0.513 |
| $a_3$ |  | 0.191 | 0.330 |  |
| $b_1$ | -0.111 | 0.007 | 0.018 | 0.210 |
| $b_2$ | 0.961 | 1.041 | 0.916 | 1.633 |
| $b_3$ |  | 2.148 | 1.925 |  |
| $c_1$ | 0.606 | 0.637 | 0.618 | 0.817 |
| $c_2$ | 0.713 | 0.584 | 0.607 | 2.279 |
| $c_3$ |  | 1.634 | 1.864 |  |

**Table S4.** NGP fitting parameters (eq S21).

## Supplementary References